\documentstyle[twoside,epsfig]{adacta}
\begin{document}
\headings{323}{333}
\title{QUANTUM CARPETS MADE SIMPLE} 
\author{I.~Marzoli, F.~Saif, I.~Bialynicki-Birula$^*$, O.~M.~Friesch, 
        A.~E.~Kaplan$^\dagger$, 
        W.~P.~Schleich\footnote{\email{schleich@physik.uni-ulm.de}}}
        {Abteilung f\"{u}r Quantenphysik, Universit\"{a}t Ulm,
        D--89069 Ulm, Germany\\
        $^*$Center for Theoretical Physics, 
        Lotnik\'ow 46, 02--668 Warsaw, Poland\\
        $^\dagger$Electr. \& Comp. Eng. Dept., The Johns Hopkins
        University, MD--21210, USA} 
%%%%%%%%%%%%%%%%%%%%%%%%%%%%%%%%%%%%%%%%%%%%%%%%%%%%%%%%%%%%%%%%%%%%
\abstract{We show that the concept of degeneracy is the key idea
          for understanding the quantum carpet woven by 
          a particle in the box.}
\def\autor{I.~Marzoli et al.}
\def\nazov{Quantum Carpets made simple}
\prvastrana=1
\poslednastrana=12
%%%%%%%%%%%%%%%%%%%%%%%%%%%%%%%%%%%%%%%%%%%%%%%%%%%%%%%%%%%%%%%%%%%%
\section{Introduction}
\begin{figure}[htb]
\begin{tabular}{cc}
       \epsfig{file=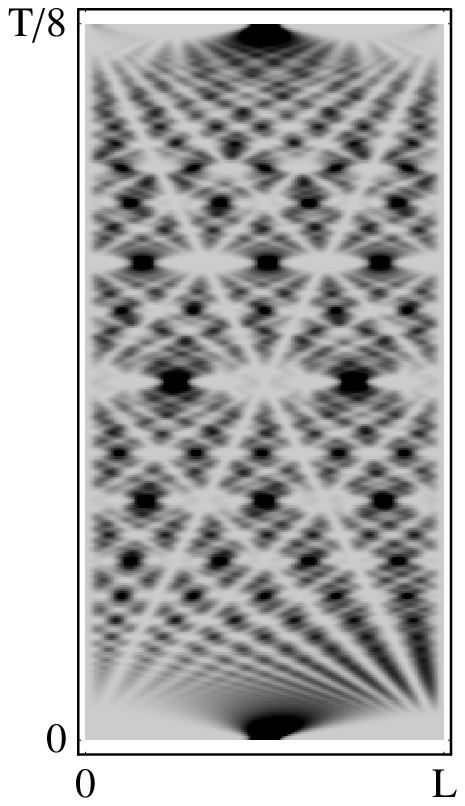, height=10truecm} & 
       \epsfig{file=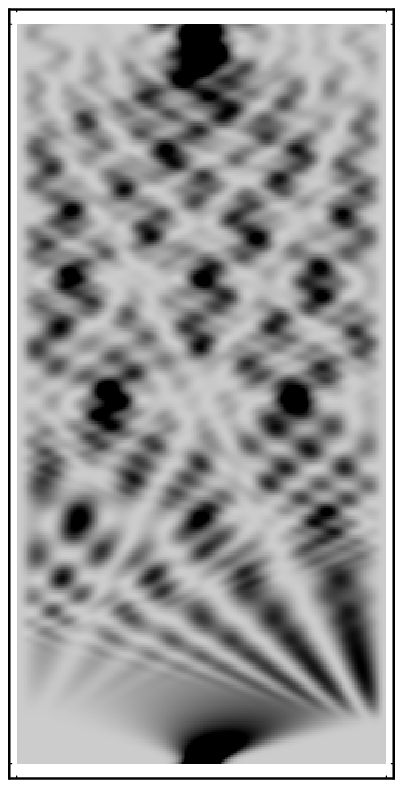, height=10truecm}
\end{tabular}
\caption{Figure 1.
         Quantum carpet woven by a non-relativistic (left) 
         and a slightly relativistic (right) particle moving
         in a one-dimensional box.
         The carpet arises from the space-time representation of 
         the probability density.
         Dark areas correspond to large probability whereas light
         areas represent low probability.
         We propagate a Gaussian wave packet according to 
         the Schr\"odinger equation with the
         non-relativistic Hamiltonian $H_{\rm nr} = p^2 / (2M)$
         or with the Hamiltonian
         $H_{\rm r} \equiv H_{\rm nr} [1 - H_{\rm nr}/(2Mc^2)]$
         approximating the relativistic Hamiltonian.
         In the latter case the straight canals and ridges of the
         non-relativistic box problem are curved.
         In both cases we use a Gaussian wave packet of width $\Delta
         x = 0.03\,L$ and average wave number $\overline{k}=10/L$
         located at $\overline{x}=L/2$.
         In the slightly relativistic example the ratio of the ground 
         state energy of the non-relativistic box and the rest mass
         is $q \equiv 10^{-6}$.
\label{fig1}}
\end{figure}

Interesting structures \cite{kinzel,berry} emerge in the 
space-time representation 
of the probability distribution for a particle in the box, as shown in
Fig.~\ref{fig1}.
Three explanations of these quantum carpets offer themselves:
Interference terms in the Wigner function \cite{stifter,marz}, 
degeneracy of intermode traces \cite{kaplan,marzoli}
and cancelation between appropriate terms of the energy representation
\cite{berry,grossmann,marklof}.
All of these explanations are rather involved.
We therefore in the present paper develop a simple argument for this 
surprising phenomenon.

We identify three properties of the particle in the box as the thread of 
the quantum carpets:
\begin{itemize}
\item[(i)] The probability density involves the product of two
standing waves creating contributions with the sum and the difference
of the wave numbers.
\item[(ii)]
The quadratic dispersion relation connecting the energy and the
momentum gives rise to a multi-degeneracy.
\item[(iii)] 
The appropriate initial conditions enhance this degeneracy.
\end{itemize}

The paper is organized as follows.
In Sect.~\ref{2} we briefly review the important formulas of the
problem of the particle in the box.  
We then in Sect.~\ref{3} give a heuristic argument 
for the quantum structures.
In the Appendix~A we derive a summation formula which allows us in 
Sect.~\ref{4} to cast
the probability density into a form which brings out most clearly 
the canals and ridges of the quantum carpets.
We conclude by summarizing the main results in Sect.~\ref{5}.
\section{The particle in the box: Fundamentals}
\label{2}
In the present section we summarize the essential ingredients of the
problem of the particle in the box.
In particular, we focus on the energy representation of the wave
function.

The probability amplitude $\psi (x,t)$ to find the particle of mass $M$ 
at time $t$ at the position $x$ in the box of length $L$ reads
\begin{equation}
\psi (x,t) = \sum_{m=1}^\infty \psi_{m} u_m (x) 
\exp\left(- \frac{i}{\hbar} E_m t\right).
\label{amplitude}
\end{equation}
Here the quantities
\begin{equation}
\psi_m \equiv \int_0^L dx \, \varphi(x) \, u_m(x)
\label{exp_coeff}
\end{equation}
are the expansion coefficients of the initial wave packet 
$\psi(x,t=0) \equiv \varphi(x)$
into the energy wave functions \cite{bohm}
\begin{equation}
u_m (x) \equiv 
  \sqrt \frac{1}{2L} \frac{1}{i} \, \left(e^{ik_mx} - e^{-ik_mx}
\right)
\label{eigenfunction}
\end{equation}
with wave numbers 
\begin{equation}
k_m \equiv m k_1 \equiv m \frac{\pi}{L}
\label{wave_number}
\end{equation}
and eigen energies
\begin{equation}
E_m \equiv \frac{(\hbar k_m)^2}{2M} = m^2 E_1 = m^2 \hbar \omega_1 
\equiv m^2\hbar \frac{2\pi}{T} .   \label{eigenenergy}
\end{equation}
In the last step we have introduced \cite{stifter1,aronstein} 
the revival time 
\begin{equation}
T \equiv \frac{4ML^2}{\hbar\pi} 
\end{equation}
at which the wave function is identical to its initial form at $t=0$,
that is $\psi(x,t=T) = \psi(x,t=0)$.

We conclude this summary of the important equations by deriving a 
representation of the probability amplitude $\psi$ that is slightly 
different from Eq.~(\ref{amplitude}).
In Sect.~\ref{4} we will use this expression to bring out the location
and shape of the structures in the quantum carpets.

We substitute the energy wave function $u_m$ into Eq.~(\ref{amplitude})
and use the expressions Eqs.~(\ref{wave_number}) and (\ref{eigenenergy})
for the wave numbers $k_m$ and energies $E_m$.
We then arrive at
\begin{equation}
\psi(x,t) = \frac{1}{i\sqrt{2L}} \sum_{m=1}^{\infty} \left\{
\psi_m \exp\left[ i\pi  m \left( \frac{x}{L}
- m\frac{2t}{T} \right) \right]
-  \psi_m \exp\left[ -i\pi  m \left( \frac{x}{L}
+ m\frac{2t}{T} \right) \right] \right\} \,.
\end{equation}
When we define the expansion coefficients $\psi_m$ for negative
values of $m$ by
\begin{equation}
\psi_{-|m|} \equiv - \psi_{|m|}
\end{equation}
we find the compact representation
\begin{equation}
\psi(x,t) = \frac{1}{i\sqrt{2L}} \sum_{m=-\infty}^{\infty} \psi_m
\exp\left[ i\pi  m \left( \frac{x}{L} - m \frac{2t}{T}
\right) \right]
\label{compact}
\end{equation}
of the wave function.
\section{Quantum carpets: A heuristic argument}
\label{3}
In the present section we use the expression Eq.~(\ref{amplitude}) 
to show that the probability density consists of four terms: Two
terms correspond to the classical trajectories, whereas the other 
two represent the striking canals and ridges of Fig.~\ref{fig1}.

Since the structures appear in the probability density
\begin{equation}
W(x,t) \equiv \psi^{\ast}(x,t) \, \psi(x,t)
\end{equation}
we now use the energy representation Eq.~(\ref{amplitude}) of $\psi$ to 
find $W$ and arrive at
\begin{equation}
W(x,t) 
      = - \frac{1}{2L} \sum_{m,n =1}^\infty \psi_m
^{\ast} \psi_n \, (e^{ik_m x} - e^{-ik_m x})(e^{ik_n x} - e^{-ik_n x})
\, \exp\left[i(k_m^2 - k_n^2)\frac{\hbar t}{2M}\right].
\end{equation}
Here we have used the expressions, Eqs.~(\ref{eigenfunction}) and 
(\ref{eigenenergy}), for the energy wave 
functions and  the energies.

When we multiply out the individual waves in the product of the two
energy wave functions
we recognize that the probability
\begin{equation}
W = I_{qc}^{(+)} + I_{qc}^{(-)} + I_{cl}^{(+)} + I_{cl}^{(-)}
\end{equation}
consists of four contributions. The terms 
\begin{equation}
I_{qc}^{(\pm)} (x,t) \equiv - \frac{1}{2 L} 
\sum_{m,n=1}^\infty \psi_m ^{\ast} \psi_n \exp \left\{ \pm i (k_m + k_n)
\left[x \pm (k_m - k_n) \frac{\hbar t}{2M} \right] \right\}
\label{qc}
\end{equation}
arise from the multiplication of the two co-propagating waves $\exp(\pm
ik_m x)$ and $\exp (\pm ik_n x)$ in the two energy wave functions. 
Note that the relation
\begin{equation}
k_m^2 - k_n^2 = (k_m + k_n)(k_m - k_n)
\label{factorization}
\end{equation}
has allowed us to factor out the sum $k_m + k_n$ of the wave numbers
in the expression Eq.~(\ref{qc}).
This creates the difference $k_m - k_n$ of the wave numbers in the
expression in the square brackets.

In contrast the terms 
\begin{equation}
I_{cl}^{(\pm)} (x,t) \equiv \frac{1}{2 L} 
\sum_{m,n=1}^\infty \psi_m ^{\ast} \psi_n \exp \left\{ \pm i (k_m - k_n)
\left[x \pm (k_m + k_n) \frac{\hbar t}{2M}\right]  \right\}
\label{cl}
\end{equation}
are a consequence of the multiplication of the two counter-propagating 
waves
$\exp (\pm ik_m x)$ and $\exp (\mp ik_n x)$ in the two energy 
wave functions. 
Here the factorization property Eq.~(\ref{factorization}) has led 
to the difference $k_m - k_n$ of the wave numbers outside the square
brackets and the sum $k_m + k_n$ inside. 

The phases 
\begin{equation}
\phi_{m,n}^{(\pm)} (x,t) \equiv \frac{x}{L} \pm  (k_m - k_n) \, 
\frac{\hbar t}{2ML} 
= \frac{x}{L} \pm (m-n) \frac{t}{T/2}
\end{equation}
and 
\begin{equation}
\Phi_{m,n}^{(\pm)} (x,t) \equiv \frac{x}{L} \pm  (k_m + k_n) \, 
\frac{\hbar t}{2ML}
= \frac{x}{L} \pm (m+n) \frac{t}{T/2}
\end{equation}
in the square brackets of Eqs.~(\ref{qc}) and (\ref{cl})
correspond to straight lines in space-time.
The steepness of these world lines is determined by the difference
and the sum of the quantum numbers $m$ and $n$.
As it was shown in Ref.~\cite{kaplan}, this opens the possibility for 
a multi-degeneracy:
Different pairs of quantum numbers $m$ and $n$ can give rise to the
same difference $m-n$ or sum $m + n$.
Therefore many world lines
can lie on top of each other enhancing in this way the contrast of the
structures.

It is the expansion coefficients $\psi_m$ that decide the question of
enhancement or suppression of these world lines.
To understand this we consider
a distribution $\psi_m$ of energy excitations
which satisfies the factorization property
\begin{equation}
\psi^*_m \, \psi_n = \psi^{(+)}_{m+n} \, \psi^{(-)}_{m-n} .
\label{factor2}
\end{equation}
Hence we can replace the product $\psi_m^* \psi_n$
of the initial expansion coefficients by another product of new 
expansions coefficients 
$\psi_s^{(+)}$ and $\psi_r^{(-)}$ which now only depend on the sum $s$
and the difference $r$ of the quantum numbers.
Any Gaussian wave packet satisfies this condition.

We conclude this section by considering a Gaussian wave packet
centered at quantum number $\overline{m}$ and width $\Delta m$ such that
$1 \ll \Delta m \ll \overline{m}$.
In this case we find a clear separation of classical and quantum 
trajectories contributing to the probability density \cite{kaplan}:
The terms $I_{cl}^{(\pm)}$ contain the classical
trajectories whereas the terms $I_{qc}^{(\pm)}$ are the origin of
the carpet.
In order to bring this out we recall that the terms $I_{cl}^{(\pm)}$ 
contain the phase 
$\Phi_{m,n}^{(\pm)}$ with the sum of the quantum numbers.
Due to the Gaussian weight factor with its maximum at $\overline{m}\gg1$
this gives rise to large quantum numbers 
of the order of $2\overline{m}$.  
The steepness of the corresponding world lines is therefore 
proportional to $\overline{m}^{-1} \ll 1$.
Consequently the world lines are rather flat and correspond to the
classical trajectories.
In contrast the terms $I_{qc}^{(\pm)}$ contain the phase 
$\phi_{m,n}^{(\pm)}$ with the difference of the quantum numbers.
This corresponds to steep world lines---the striking canals and ridges.

\section{A new representation of the probability density}
\label{4}
In the preceding section we have shown 
that for the evaluation of $I_{qc}^{(\pm)}$ and $I_{cl}^{(\pm)}$
it is natural to introduce the new summation indices $m \pm n$.
In the present section we pursue this idea.
However, we do not start from Eqs.~(\ref{qc}) and (\ref{cl}) but from
Eq.~(\ref{compact}).

The probability $W$ then reads
\begin{equation}
W(x,t) = \frac{1}{2L} \sum_{m,n=-\infty}^{\infty} \psi_m^* \psi_n
\, \exp\left\{-i\pi(m-n) \left[ \frac{x}{L} - (m+n) \frac{2t}{T}
\right] \right\} .
\end{equation}
With the help of the summation formula
\begin{equation}
\sum_{m,n = -\infty}^\infty f_{m,n} = \frac{1}{2}
  \sum_{l = -\infty}^\infty
  \sum_{j = -\infty}^ \infty (-1)^{jl}
  \int_{-\infty}^\infty d\rho \, f\left[\frac{1}{2} (j +\rho),
 \frac{1}{2} (j-\rho)\right] \, \exp (i \pi l \rho) 
\end{equation}
derived in the Appendix~A the probability density takes the form
\begin{equation}
W(x,t) = \frac{1}{4L} \sum_{l,j=-\infty}^{\infty} (-1)^{jl} \,
\Psi^{(W)} \! \left[ \frac{j}{2}, \, \chi_{j,l}(x,t) \right] .
\label{density}
\end{equation}
Here we have introduced the Wigner function 
\begin{equation}
\Psi^{(W)} (\mu,\xi) \equiv 
\int_{-\infty}^\infty d\rho \, \psi^* \! \left[\mu+\frac{\rho}{2}\right]
\, \psi \! \left[\mu-\frac{\rho}{2}\right] \, \exp( -i\pi \rho \xi),
\label{wigner}
\end{equation}
of the expansion coefficients $\psi_m$.
Note that $\psi[\mu]$ is a continuous extension of $\psi_m$
such that $\psi[\mu] \equiv \psi_m$ for $\mu = m$.

Moreover, we have defined 
\begin{equation}
\chi_{j,l}(x,t) \equiv \frac{x}{L} - j \frac{t}{T/2} -l.
\end{equation}

Expression Eq.~(\ref{density}) for the probability density is the main
result of the present paper.
It brings out most clearly that the probability density $|\psi(x,t)|^2$
consists of a superposition of structures $\Psi^{(W)}$ aligned along 
straight lines defined by $\chi_{j,l}(x,t)$.
The Wigner function $\Psi^{(W)}$ of the expansion coefficients 
determines the shape of these structures.

We conclude this section by making contact with the expression,
\begin{equation}
W(x,t) = \frac{\pi\hbar}{2L} \sum_{l,j=-\infty}^{\infty} (-1)^{jl} \,
\Phi^{(W)} [ \chi_{j,l}(x,t), \, p_j ] ,
\label{density2}
\end{equation}
derived in Ref.~\cite{stifter}.
Here $p_j \equiv j \pi \hbar/(2L)$ and
\begin{equation}
\Phi^{(W)}(x,p) \equiv 
\frac{1}{2\pi\hbar} \int_{-\infty}^{\infty} dy \,
\phi \left( x + \frac{y}{2} \right) \, 
\phi^* \left( x - \frac{y}{2} \right) \, \exp (-i p y/\hbar)
\end{equation}
is the Wigner function \cite{hillery} of the superposition wave function
\begin{equation}
\phi(x) \equiv \varphi(x) - \varphi(-x)
\end{equation}
built out of the original wave packet and its mirror image.

The similarity of the two expressions Eqs.~(\ref{density}) and 
(\ref{density2})
for the probability density already suggests that the Wigner function
$\Psi^{(W)}$ in $(\mu,\xi)$ space is related to the Wigner function
$\Phi^{(W)}$ in $(x,p)$ phase space.
Indeed in Appendix~B we derive the relation
\begin{equation}
\Psi^{(W)}(\mu,\xi) = 2\pi \hbar \, \Phi^{(W)} \left( L\xi, 
\frac{\pi\hbar}{L} \mu\right) . 
\end{equation}
Hence the two arguments $\mu$ and $\xi$ appropriately scaled now
play the role of momentum $p\equiv \hbar k_\mu$ and position
$x \equiv L\xi$.
\section{Conclusions}
\label{5}
In the energy representation the probability of finding the particle
at time $t$ at position $x$ is a double sum over all energy quantum 
numbers.
The quadratic dispersion relation of the free particle 
allows us to express this double sum by another double sum 
extending over the sums and differences of the quantum
numbers. 
In this way we represent the probability distribution in space-time 
as a superposition of structures along straight world lines. 
Their steepness and their starting points are determined
by integers.
The shape of the initial wave packet, that is the form $\psi_m$ of the
initial excitation of the energy eigenstates governs the shape of these 
structures.

We emphasize that the expression Eq.~(\ref{density}) for the probability
density derived in the present paper
is very different from the one which follows from the representation
\cite{berry,stifter1,aronstein}
\begin{equation}
\psi\left(x,t =\frac{q}{r}T+\Delta t\right) = \sum_{l=-\infty}^{\infty}
{\cal W}_l^{(r)} \, \varphi \! \left(x - \frac{l}{r}2L, \Delta t \right)
- \sum_{l=-\infty}^{\infty} {\cal W}_l^{(r)} \, 
\varphi \! \left(-x+ \frac{l}{r}2L, \Delta t \right),
\label{rev}
\end{equation}
of the wave function $\psi$ in the neighborhood of a fraction $q/r$
of the revival time $T$.
Here 
\begin{equation}
{\cal W}_l^{(r)} \equiv \frac{1}{r} \sum_{p=0}^{r-1} \exp \left[
-2\pi i \left( p^2 \frac{q}{r} - p \frac{l}{r} \right) \right]
\end{equation}
denotes the Gauss sums \cite{lang,hannay} and
\begin{equation}
\varphi (x,t) \equiv \int_0^L dx' \, G_{free}(x,t|x',0)
\, \varphi(x')
\end{equation}
is the initial wave function $\varphi(x)\equiv \psi(x,t=0)$ propagated
freely according to the Green's function $G_{free}$ of the free particle.

Indeed the above formula Eq.~(\ref{rev})
is a local representation in space-time
whereas Eq.~(\ref{density}) is a global one:
It depicts the probability density $W$ as a
superposition of structures along {\em vertical\/} and {\em tilted\/}
world lines, whereas the revival representation Eq.~(\ref{rev})
uses a superposition of 
structures along lines of constant time, that is along horizontal world
lines.

We have obtained these results using
the summation formula derived in the Appendix~A.
This formula has a much wider range of application.
For example, it immediately provides an expression for the slightly 
relativistic particle.
In this case the structures are not along straight but curved lines
as shown in Fig.~\ref{fig1} (right).
Moreover, it provides insight into the development of fractal canals
discovered in Ref.~\cite{berry}, when the initial wave packet is 
uniform.
However, space does not allow us to go deeper into these topics of 
future publications. 
%%%%%%%%%%%%%%%%%%%%%%%%%%%%%%%%%%%%%%%%%%%%%%%%%%%%%%%%%%%%%%%%%%%%%%  
\section*{Appendix A: A useful summation formula} 
In this appendix we derive two different but equivalent representations 
of the double sum
\begin{equation}
I \equiv \sum_{m,n = -\infty}^{\infty} f_{m,n}
\end{equation}
with coefficients $f_{m,n}$.

Our derivation relies on the introduction of the new summation indices 
$m+n \equiv s$ and $m-n \equiv r$.
Note however, that this definition puts certain restrictions 
on $s$ and $r$. 
Indeed, we have to distinguish two cases: (i) when $m$ and $n$
are both even or odd, and (ii) when one of them is odd and the 
other is even. 
In the case (i) we find that $m-n$ and $m+n$ are both even. 
Hence we have the substitutions
\begin{equation}
m-n \equiv 2 r \quad \mbox{and} \quad m+n \equiv 2 s.
\end{equation}
In the case (ii) we find that $m-n$ and $m+n$ are both odd, 
which leads to the definition
\begin{equation}
m-n \equiv 2 r + 1 \quad \mbox{and} \quad m+n \equiv 2s + 1.
\end{equation}
We therefore find the rule
\begin{equation}
\sum_{m,n = - \infty}^ \infty f_{m,n} 
= \sum_{r,s = - \infty}^ \infty f_{s+r, s-r} 
+ \sum_{r,s = - \infty}^ \infty f_{s+r+1, s-r}
\label{sr}
\end{equation}
for replacing the original sums 
extending over $m$ and $n$ by new sums extending over $r$ and $s$.

We can combine the two terms in Eq.~(\ref{sr}) into one, 
when we replace either the summation over $r$, or the one over $s$,
by an integration.
The Poisson summation formula 
\cite{lighthill}
\begin{equation}
\sum_{m = -\infty}^\infty g_m = \sum_{l = -\infty}^\infty 
\int_{-\infty}^\infty d \mu \, g[\mu] \, \exp(2\pi i l\mu)
\label{poisson}
\end{equation}
allows us to do this in an exact way.
Here $g [\mu]$ is an extension of the function $g_m$ to the whole real 
axis such that $g[\mu]$ takes on the values $g_m$ at integer values
$\mu =m $.

When we apply the Poisson formula to the summation over $r$ we 
arrive at 
\begin{eqnarray}
I &=& \sum_{l = -\infty}^\infty \left\{  \sum_{s = -\infty}^\infty 
\int_{-\infty}^\infty
d \rho \, f[s+\rho, s-\rho] \, \exp (2\pi i l \rho)\right.\nonumber\\
&+&\left.  \sum_{s = -\infty}^\infty \int_{-\infty}^\infty d \rho \,
f[s+\rho + 1, s-\rho] \, \exp (2 \pi i l \rho) \right\},
\end{eqnarray}
which after the substitutions $\tilde{\rho} \equiv 2\rho$ 
and $\bar{\rho} 
\equiv 2 \rho + 1$
in the two integrals takes the form
\begin{eqnarray}
I &=& \frac{1}{2}\sum_{l = -\infty}^ \infty \left\{  
\sum_{s = -\infty}^ \infty 
\int_{-\infty}^\infty d \tilde{\rho} \, f\left[ \frac{1}{2} (2s + 
\tilde{\rho}),
\frac{1}{2} (2s-\tilde{\rho}) \right] \, \exp(i\pi l\tilde{\rho})\right. 
\nonumber\\
&+&\left.  \sum_{s = -\infty}^ \infty (-1)^l \int_{-\infty}^\infty d 
\bar{\rho} \, f \left[\frac{1}{2} (2s +1+ \bar{\rho}), 
\frac{1}{2} (2s +1- \bar{\rho}) \right]  \,
\exp (i \pi l \bar{\rho})\right\}.
\end{eqnarray}
Here we have used for the last integral the relation 
\begin{equation}
\exp (i\pi l) = (-1)^l ,
\end{equation}
and have written the arguments of the function $f$ in a way that brings 
out most clearly that the two integrals are the even and odd
terms of a single sum. The
only obstacle left before we can combine these two terms 
is the term $(-1)^l$. 
When we recall that 
\begin{equation}
    (-1)^{jl}  
    =  \left\{ \begin{array}{ll} 
                   (-1)^{2sl}=1          & \mbox{for } j=2s \\ 
                   (-1)^{(2s+1)l}=(-1)^l & \mbox{for } j=2s+1
               \end{array}   
       \right .
\end{equation}
we find indeed
\begin{equation}
\sum_{m,n = -\infty}^\infty f_{m,n} = \frac{1}{2}
\sum_{l = -\infty}^\infty 
\sum_{j = -\infty}^ \infty (-1)^{jl}  
\int_{-\infty}^\infty d\rho \, f\left[\frac{1}{2} (j +\rho),
\frac{1}{2} (j-\rho)\right] \, \exp (i \pi l \rho) .
\label{I1}
\end{equation}
We conclude this appendix by presenting a different expression for the
double sum $I$ which follows from Eq.~(\ref{sr}) when we replace
the summation over $s$ by an integration using the
Poisson summation formula, Eq.~(\ref{poisson}).
In this case we find following the same train of thought 
\begin{equation}
\sum_{m,n = -\infty}^\infty f_{m,n} = \frac{1}{2}
\sum_{l = -\infty}^ \infty 
\sum_{j = -\infty}^ \infty (-1)^{jl}  
\int_{-\infty}^\infty d\sigma \, f \left[\frac{1}{2} (\sigma+j),
\frac{1}{2} (\sigma-j)\right] \, \exp (i \pi l \sigma) .
\label{I2}
\end{equation}

We recognize that the two representations are different:
In the one of Eq.~(\ref{I1}) the integration variable $\rho$
enters in an asymmetric way whereas in the one of Eq.~(\ref{I2})
the integration variable $\sigma$ appears in a symmetric way.
Nevertheless, both representations are completely equivalent.
\section*{Appendix B: Relation between the two Wigner functions}
In this appendix we relate the Wigner function
\begin{equation}
\Psi^{(W)} (\mu,\xi) \equiv
\int_{-\infty}^\infty d\rho \, \psi^* \! \left[\mu+\frac{\rho}{2}\right]
\, \psi \! \left[\mu-\frac{\rho}{2}\right] \, \exp( -i\pi \rho \xi)
\label{b1}
\end{equation}
of the continuous extension $\psi[\mu]$ of the expansion coefficients 
\begin{equation}
\psi_m \equiv \int_0^L dx \, \varphi(x) \, u_m(x)
\end{equation}
to the Wigner function
\begin{equation}
\Phi^{(W)}(x,p) \equiv \frac{1}{2\pi\hbar} \int_{-\infty}^{\infty} dy \,
\phi\left( x+\frac{y}{2} \right) \, 
\phi^*\left( x- \frac{y}{2} \right) 
\, e^{-i p y / \hbar}
\label{b2}
\end{equation}
in position $x$ and momentum $p$ of the superposition state
\begin{equation}
\phi(x) \equiv \varphi(x) - \varphi(-x) .
\end{equation}

For this purpose we first note that the initial wave packet vanishes at 
the walls at $x=0$ and $x=L$.
Moreover, it vanishes outside of the box.
We can therefore extend the integral in the definition of the expansion
coefficients to $-\infty$ and $+\infty$, that is
\begin{equation}
\psi_m \equiv \sqrt{\frac{2}{L}} \int_{-\infty}^{\infty} dx \,\varphi(x) 
\, \sin\left( m\pi\frac{x}{L}\right)
\end{equation}
where we have used the definition Eq.~(\ref{eigenfunction}) 
of the energy eigenfunctions.

When we substitute this expression into the definition of the Wigner
function Eq.~(\ref{b1}) we arrive at
\begin{eqnarray}
\Psi^{(W)} (\mu, \xi) &=& - \frac{1}{2L} \int_{-\infty}^{\infty} dx'
\int_{-\infty}^{\infty} dx'' \, \varphi^*(x') \, \varphi(x'')
\nonumber \\  
&\times & \left\{ \exp\left( i\pi\mu\frac{x'+x''}{L} \right)
         \int_{-\infty}^{\infty} d\rho \, \exp\left[ i\pi\rho \left(
         \frac{x'-x''}{2L} - \xi \right) \right] 
\right. \nonumber \\
&-&   \exp\left( i\pi\mu\frac{x'-x''}{L} \right) 
      \int_{-\infty}^{\infty} d\rho \, \exp\left[ i\pi\rho \left(
      \frac{x'+x''}{2L} - \xi \right) \right] \nonumber \\
&-&   \exp\left(-i\pi\mu\frac{x'-x''}{L} \right)
      \int_{-\infty}^{\infty} d\rho \, \exp\left[ i\pi\rho \left(
      - \frac{x'+x''}{2L} - \xi \right) \right] \nonumber \\
&+& \left. \exp\left(-i\pi\mu\frac{x'+x''}{L} \right)
          \int_{-\infty}^{\infty} d\rho \, \exp\left[ i\pi\rho \left(
          -\frac{x'-x''}{2L} - \xi \right) \right]
\right\} .
\end{eqnarray}

We now introduce the new integration variables 
$\overline{x}'' \equiv -x''$
in the second, $\overline{x}' \equiv - x'$ in the third and the pair
$\overline{x}' \equiv -x'$ and $\overline{x}'' \equiv -x''$ in the
forth term of the brackets.
This allows us to combine the four contributions which yields
\begin{eqnarray}
\Psi^{(W)} (\mu, \xi) &=& - \frac{1}{L} \int_{-\infty}^{\infty} 
d\overline{x}' \int_{-\infty}^{\infty} d\overline{x}'' \,
\left[ \varphi^*(\overline{x}') - \varphi^*(-\overline{x}') \right]
\left[ \varphi(\overline{x}'') - \varphi(-\overline{x}'') \right]
\nonumber \\
& \times & \exp \left( i\pi\mu \frac{\overline{x}'+\overline{x}''}{L}
\right) \, \delta \! \left( \frac{\overline{x}' - \overline{x}''}{2L}
-\xi \right) .
\end{eqnarray}
Here we have also made use of the relation
\begin{equation}
\int_{-\infty}^{\infty} d\rho \, e^{i\pi\rho \zeta} = 2 \, \delta(\zeta).
\end{equation}
When we now perform the integration with the help of the delta function
we find
\begin{equation}
\Psi^{(W)} (\mu, \xi) = - 2 \int_{-\infty}^{\infty} dx'' \,
\phi^*(x''+2L\xi) \, \phi(x'') \, \exp\left(2\pi i \mu 
\frac{x''+L\xi}{L} \right)
\end{equation}
which with the new integration variable $-y/2 \equiv x''+L\xi$ reads
\begin{equation}
\Psi^{(W)} (\mu,\xi) = - \int_{-\infty}^{\infty} dy \, 
\phi^* \left(L\xi - \frac{y}{2} \right) \, 
\phi\left(-L\xi - \frac{y}{2} \right) \, \exp(-i\pi\mu y/ L)
\end{equation}
or 
\begin{equation}
\Psi^{(W)} (\mu,\xi) = \int_{-\infty}^{\infty} dy \,
\phi^* \left(L\xi - \frac{y}{2} \right) \,
\phi \left(L\xi + \frac{y}{2} \right) \, \exp(-i\pi\mu y/L)
\end{equation}
where we have made use of the symmetry relation $\phi(-x) = - \phi(x)$.

When we compare this expression to the Wigner function $\Phi^{(W)}(x,p)$,
Eq.~(\ref{b2}), in position $x$ and in momentum $p$ we find the relation
\begin{equation}
\Psi^{(W)} (\mu,\xi) = 2 \pi \hbar \, \Phi^{(W)} \left(L\xi, 
\frac{\pi\hbar}{L} \mu \right).
\end{equation}

We conclude this appendix by noting that the operation of complex 
conjugation enters differently in the two Wigner functions: 
In $(x,p)$ phase space we take the complex conjugate of the wave function
with the argument $x-y/2$.
In contrast, in the Wigner function in $(\mu,\xi)$ space we take
the complex conjugate of the expansion coefficient with the argument
$\mu + \rho /2$.
\section*{Acknowledgement}
We express our gratitude to P.~J.~Bardroff, M.~V.~Berry, J.~H.~Eberly,
M.~Fontenelle, F.~Gro{\ss}mann,
M.~Hall, H.~J.~Kimble, T.~Kiss, W.~E.~Lamb, Jr.,
K.~A.~H. van Leeuwen, C.~Leichtle, J.~Marklof,
M.~M.~Nieto, J.~M.~Rost, E.~C.~G.~Sudarshan
and P.~Stifter for many fruitful
discussions on this topic.
Two of us (I.~B.-B.) and (A.~E.~K.) thank the Humboldt Stiftung for
their support.

\vspace{1truecm}

%
%References

%%

\begin{thebibliography}{99}
%\bibitem{lab1}\refer{P.W. Anderson}{Phys. Rev. B}{120}{1959}{12345}
%\bibitem{lab2} A.B. Author: {\sl name of book}, Publishing House, year.
\bibitem{kinzel} \refer{W.~Kinzel}{Phys. Bl.}{51}{1995}{1190}
\bibitem{berry} \refer{M.~V.~Berry}{J. Phys. A}{29}{1996}{6617}
        \refer{M.~V.~Berry, S. Klein}{J. Mod. Optics}{43}{1996}{2139}
\bibitem{stifter} \refer{P.~Stifter, C.~Leichtle, W.~P.~Schleich,
         J. Marklof}{Z. Naturf.}{52 a}{1997}{377}
\bibitem{marz} I.~Marzoli, O.~M.~Friesch, W.~P.~Schleich: {\sl 
         Proceedings of the Fifth Wigner Symposium\/}, 
         ed. P.~Kasperkovitz, World Scientific, in press;
\bibitem{kaplan} \refer{A.~E.~Kaplan, P.~Stifter, K.~A.~H. van Leeuwen,
W.~E.~Lamb,Jr., W.~P.~Schleich}{Physica Scripta}{}{in press}{}
\bibitem{marzoli} I.~Marzoli, I.~Bialynicki-Birula, O.~M.~Friesch,
         A.~E.~Kaplan, W.~P.~Schleich: {\sl Proceedings of the
         Physical Research Laboratory Jubilee Conference on 
         Nonlinear Dynamics and Computational Physics\/}, 
         ed. V.~B.~Sheorey, Narosa Publishing House, in press
         [see also {\sl Los-Alamos e-print archive\/} quant-ph/9804015
         (1998)];
\bibitem{grossmann} \refer{F.~Gro{\ss}mann, J.-M. Rost, 
         W.~P.~Schleich}{J. Phys. A}{30}{1997}{L277}
\bibitem{marklof} J. Marklof: {\sl Limit theorems for theta sums with 
        applications in quantum mechanics\/} 
        Shaker Verlag, 1997;
\bibitem{bohm} See, e.g. D.~Bohm: {\sl Quantum theory\/}, 
         Prentice-Hall, 1951;
\bibitem{stifter1} P.~Stifter, W.~E.~Lamb,Jr., W.~P.~Schleich:
         {\sl Proceedings of the Conference on Quantum Optics and Laser
         Physics\/}, ed. L.~Jin and Y.~S.~Zhu, World Scientific, 1997;
         I.~Marzoli, O.~M.~Friesch, P.~Stifter, W.~P.~Schleich:
         to be published;
\bibitem{aronstein} \refer{D.~L.~Aronstein, C.~R.~Stroud,Jr.}{Phys. 
         Rev. A}{55}{1997}{4526}
\bibitem{hillery} \refer{M.~Hillery, R.~F.~O'Connell, M.~O.~Scully,
         E.~P.~Wigner}{Phys. Rep.}{106}{1984}{121}
\bibitem{lang} S.~Lang: {\sl Algebraic Number Theory\/} 
         Addison-Wesley, 1970;
\bibitem{hannay} \refer{J.~H.~Hannay, M.~V.~Berry}{Physica D}{1}
         {1980}{267}
\bibitem{lighthill} M.~J.~Lighthill: {\sl Introduction to Fourier 
         analysis and generalized functions\/}, Cambridge University 
         Press, 1960.
\end{thebibliography}
\end{document}